\documentclass[twocolumn,showpacs,preprintnumbers,amsmath,amssymb,aps,prb]{revtex4-1}

\usepackage{graphicx}
\begin{document}

\title[Ferroelectric charge order stabilized by antiferromagnetism...]{Ferroelectric charge order stabilized by antiferromagnetism in multiferroic LuFe$_2$O$_4$.}

\author{A.M. Mulders$^{1}$,  M. Bartkowiak$^{1,2}$, J.R. Hester$^{2}$, E. Pomjakushina$^{3}$, K. Conder$^{3}$}

\address{$^{1}$School of Physical, Environmental and Mathematical Sciences,
UNSW@ADFA, Canberra ACT 2600, Australia}

\address{ $^{2}$The Bragg Institute, Australian Nuclear
Science and Technology Organisation, Lucas Heights, NSW 2234, Australia}

\address{ $^{3}$Laboratory for Development and Methods,
Paul Scherrer Institut, 5232 Villigen PSI, Switzerland}

\begin{abstract}
Neutron diffraction measurements  on multiferroic LuFe$_2$O$_4$ show changes in the antiferromagnetic (AFM) structure characterized by wavevector $q$ = ($\frac{1}{3} \frac{1}{3} \frac{1}{2}$) as a function of electric field cooling procedures. The increase of intensity from all magnetic domains and the decrease in the 2D magnetic order observed below the N\'{e}el temperature are indicative of increased ferroelectric charge order. The AFM order changes the dynamics of the CO state, and stabilizes it. It is determined that the increase in electric polarization observed at the magnetic ordering temperature is due to a transition from paramagnetic 2D charge order to AFM 3D charge order. 
\end{abstract}

\maketitle

\noindent Ferroelectric polarization (FE) arises from covalent bonding between anions and cations or the orbital hybridization of electrons and generally excludes occurrence of finite $3d$ magnetic moments and associated magnetism. \cite{hill_jbcb_2000} However, competing bond and charge order in transition metal compounds can lead to inversion symmetry breaking and multiferroic behaviour. \cite{vandenbrink_jpcm_2008} Alternatively, FE polarization may arise from  frustrated charge order (CO) as reported for LuFe$_2$O$_4$. \cite{ikeda_nat_2005}
This compound is of particular interest, as, in addition to
ferroelectricity, magnetism originates from the same Fe ions and this holds the promise of strong
magnetoelectric coupling. The FE and magnetic order (MO) takes place at and near ambient
temperature, respectively, which provides the potential for room temperature multiferroics.
These properties make them extremely useful in devices where the coexistence of both charge and spin
components can be exploited and one property can be used to drive the other. 

LuFe$_2$O$_4$ has generated a large scientific interest since its multiferroic properties were reported by Ikeda.\cite{ikeda_nat_2005} LuFe$_2$O$_4$ adopts a FE ground state below T$_{CO}\sim$350~K, while below T$_N\sim$240~K the Fe magnetic moments order which
enhances the FE polarization by 20\%. 
It has been demonstrated that the dielectric properties can be changed with applied magnetic field \cite{subramanian_am_2006, wen_prb_2009} and that pulsed currents can change the magnetic state,\cite{li_prb_2009, wang_jpcm_2010} although the latter is under debate.\cite{wen_prb_2010} These studies confirm the strong coupling between the magnetic and FE orders, yet the intrinsic nature of the charge order suggests the charge and magnetic order are in principle independent. \cite{vandenbrink_jpcm_2008}  For example, it has been recently demonstrated that the magnetic and charge fluctuations are largely independent 
in the frustrated magnet NiGa$_2$S$_4$. \cite{takubo_prl_2010}

\begin{figure}[t]
\includegraphics[width=0.8\columnwidth]{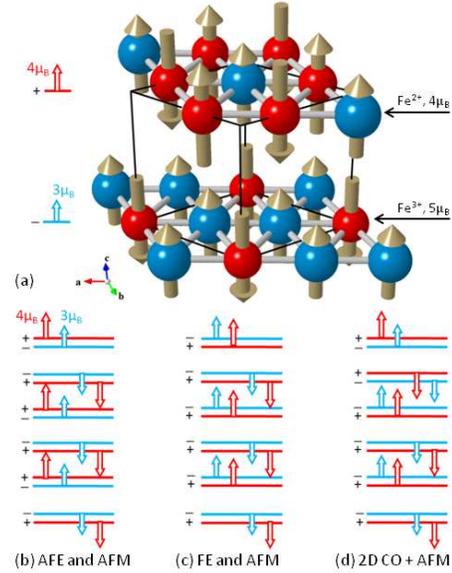}
\caption{\label{fig_layers} (a) Drawing \cite{vesta} of the FeO$_5$ bilayer indicating small red  Fe$^{3+}$ and  large blue Fe$^{2+}$ ions with magnetic moments of 5 $\mu_B$ and 4 $\mu_B$. The $\sqrt{3} \times \sqrt{3}$ unit cell is indicated in bold solid lines and the total magnetic moment per layer is indicated at the left. 
(b,c,d) AFM order for (b) AFE CO, (c) FE CO and (d) disordered CO (2D) respectively.
The n.n. distance of iron ions is 3.65 \AA~in a single layer, 3.23 \AA~in a bilayer, and 6.34 \AA~between bilayers.
}
\end{figure}

Various electron, x-ray and neutron diffraction (ND) studies have investigated the charge and magnetic orders.
The crystal structure of rhombohedral LuFe$_2$O$_4$ ($R\bar{3}m$) consists of a triangular double layer of iron ions, with trigonal bipyramids of
five oxygen nearest neighbors (n.n.), in which an equal amount of Fe$^{2+}$ and Fe$^{3+}$ are believed to coexist at the same site (Fig \ref{fig_layers}(a)).
The observation of the ($\frac{H}{3} \frac{H}{3} \frac{L}{2}$)  reflections using resonant x-ray Bragg diffraction (RXD) further supports the existence of CO,\cite{ikeda_jpcm_2008} 
and we demonstrated that the Fe$^{2+}$ orbitals exhibit a glass like order.\cite{mulders_prl_2009}

The triangular lattice gives frustration for a 1:1 ratio of  Fe$^{2+}$ to  Fe$^{3+}$ ions, but is stable for a 1:2 ratio where one species is surrounded by a 
honeycomb lattice of the other. Thus one of the layers has a majority of Fe$^{2+}$ ions and the other a majority of Fe$^{3+}$ ions, which results in n.n. of the opposite species for all minority ions. Such CO gives rise to an electric dipole moment.
While the CO state in the bilayers is understood, the sign of electrostatic interaction between the bilayers is unclear. The spontaneous polarization suggests a FE ground state, where the FE moments of the bilayers are aligned along the $c$-axis. In contrast an antiferroelectric (AFE) ground state has been suggested from x-ray diffraction.\cite{angst_prl_2008}

In the present study we investigate the magnetic order in LuFe$_2$O$_4$ with single crystal  ND as a function of applied electric field cooling procedures.
We observe a change in the magnetic long range order when an applied electric field is present at T$_N$. Our results show that long range CO that is promoted by the electric field, is stabilized by the magnetic order. We propose that AFM order in the Fe bilayers facilitates the formation of long range CO because electron hopping between the sheets of the bilayer is assisted by the alignment of the Fe$^{2+}$ and  Fe$^{3+}$ spins. 

\begin{figure}[t]
\includegraphics[width=0.75\columnwidth,angle=0]{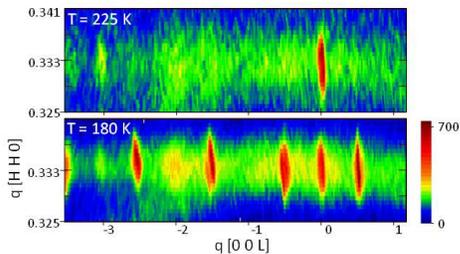}
\caption{\label{fig_images} ND intensity recorded in the [H H L] plane at (a) T=225~K and (b) T=180~K, after cooling the sample from T=350~K in zero applied electric field. At T=225~K a broad magnetic rod is observed at ($\frac{1}{3} \frac{1}{3}$ L) characteristic of 2D magnetic ordering. The peak observed at ($\frac{1}{3} \frac{1}{3}$0) is characteristic of FM alignment of the FeO$_5$ bilayers. At T=180~K magnetic superlattice peaks are observed at ($\frac{1}{3} \frac{1}{3} \frac{L}{2}$) characteristic of AFM alignment of the  FeO$_5$ bilayers.
}
\end{figure}

A LuFe$_2$O$_4$ single crystal was prepared as reported previously and cut from the same ingot as the crystals used for the RXD study. \cite{mulders_prl_2009} 
Pyroelectric current measurement shows a decrease of the electric polarization above T=220~K of $\sim$10\% and the sample becomes conductive due to the loss of CO above T$_{CO}\sim$330~K. 

ND experiments were performed at the Wombat diffractometer at the OPAL facility in Sydney, using 2.95 \AA~monochromatic neutrons. The wavelength was calibrated using Al$_2$O$_3$, and the efficiency of the area detector was calibrated with a vanadium sample. The diffraction peaks of interest were selected from the area detector to reduce the background. We noted several other crystallites besides the one of this study.
The selected single crystal was aligned with the [110] and [001] axes in the scattering plane and mounted between two vanadium electrodes 
giving rise to an electric field in the scattering plane at an angle of 30 degrees to the [110] axis. The sample was electrically isolated since thermal contact was achieved with a sapphire rod. 
The electric field was applied in vacuum without contact to the sample but surface currents or shorts through a part of the sample 
cannot be excluded. 
The electric field was removed  and thermal equilibrium was achieved before each ND measurement.

We recorded ND spectra at T=225~K and at T=180~K, following a sequence of zero electric field cool (ZEFC) and electric field cool (EFC) sequences. 
First, the sample was heated to T=350~K and cooled without electric field to obtain a ZEFC ground state. Fig. \ref{fig_images} shows the diffraction intensity along ($\frac{1}{3} \frac{1}{3}$L),  recorded at T=225~K and T=180~K respectively. The rod of intensity along ($\frac{1}{3} \frac{1}{3}$L) observed at T=225~K is characteristic of 2D magnetic order in the $ab$ planes. Similar behavior is reported by Ref. \cite{iida_jpsj_1993}. The $q$-dependence of the intensity is well described by 2D magnetic order with magnetic moments along the $c$-axis. Because of the low $q$ range of our experiment we are not sensitive to the CO. 

At T=180~K superlattice peaks are observed at half integer values of L, indicating antiferromagnetic (AFM) order, while superlattice peaks at integer L are largely absent. The reflection observed at ($\frac{1}{3} \frac{1}{3}$0) does not change as function of temperature and electric field. It is believed to 
be distinct from the AFM structure and will be discussed later. First we focus on the half integer superlattice.
In contrast with ref. \cite{christianson_prl_2008}, where the half integer peaks are attributed to ferrimagnetic alignment of Fe$^{2+}$ and Fe$^{3+}$ moments, we observe all half integer L without the presence of all integer L reflections. This suggests that AFM and ferromagnetic (FM) arrangements of the bilayers occur in distinct regions of the crystal and are possibly formed depending on the stoichiometry of the sample. 

\begin{figure}[t]
\includegraphics[width=1.2\columnwidth]{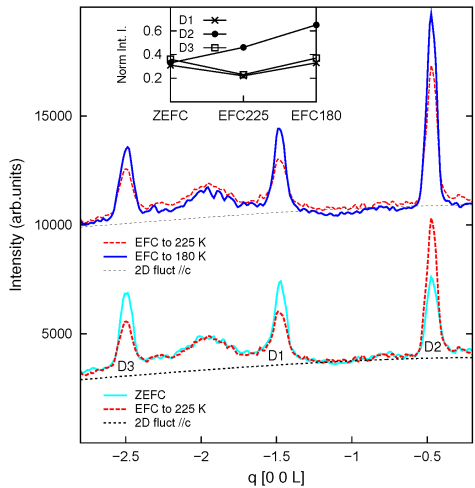}
\caption{\label{fig_L} Neutron intensity recorded along the ($\frac{1}{3} \frac{1}{3}$  L) direction at T=180~K as function of EFC procedure. 
The magnetic state obtained after EFC to T=225~K is compared to the magnetic state obtained after ZEFC (bottom) and  to the magnetic state obtained after EFC to T=180~K (top, shifted for clarity). The black dotted line indicates 2D magnetic order in the $ab$ plane with moments along the $c$ axis.
The inset shows the integrated intensity, normalized to the total intensity at ZEFC, as function of EFC sequence.  D1, D2 and D3 correspond to {\bf m}$_1$, {\bf m}$_2$ and {\bf m}$_3$ respectively. 
}
\end{figure}

Symmetry allows for three CO domains that are characterized by propagation vectors  {\bf p}$_A$=($\frac{1}{3}$, $\frac{1}{3}$, $\frac{3}{2}$),  {\bf p}$_B$=($\frac{-2}{3}$, $\frac{1}{3}$, $\frac{3}{2}$) and  {\bf p}$_C$=($\frac{1}{3}$, $\frac{-2}{3}$, $\frac{3}{2}$) \cite{angst_prl_2008} and three magnetic domains, characterized by {\bf m}$_1$= ($\frac{1}{3}$, $\frac{1}{3}$, $\frac{3}{2}$), {\bf m}$_2$= ($\frac{-2}{3}$, $\frac{1}{3}$, $\frac{3}{2}$) and {\bf m}$_3$=($\frac{1}{3}$, $\frac{-2}{3}$, $\frac{3}{2}$). Fig. \ref{fig_layers}(a)) shows a bilayer with {\bf p}$_A$ and {\bf m}$_2$. Order with {\bf p}$_A$ and {\bf m}$_3$ has the same motif, while order with {\bf p}$_A$ and {\bf m}$_1$ is different.
Distinct magnetic intensities are calculated from the structure factor,
and the similar intensity suggests a large amount of magnetic and/or lattice stacking faults along the $c$-axis (Fig. \ref{fig_L}).

Further, the sample was heated to T=350~K and cooled to T=225~K in an applied electric field of 350~kV/m. The neutron intensity recorded at T=225~K is identical to that of the ZEFC case. We emphasize that the 2D rod observed at T=225~K is magnetic while the CO is predominantly 3D, as observed with RXD.\cite{mulders_prl_2009} 

The electric field was removed before the ND was recorded, and before the sample was further cooled to T=180~K. Fig \ref{fig_L} compares the ND intensity recorded at T=180~K with the ND intensity recorded after ZEFC. The intensity of domain D2 has increased significantly while the intensity of domains D1 and D3 has decreased, indicating a change in magnetic domain population after EFC to T=225~K. 
The width of the superlattice peaks along [110] is limited by the resolution of the instrument while it is larger than the resolution along [001] and well described with Gaussian peak profiles. The magnetic correlation length is taken as the inverse of the half width at half maximum. The fitted integrated intensities and correlation lengths are summarized for the three domains in the inset of Fig. \ref{fig_L} and in Table \ref{table}. 

The correlation length of the majority domain after EFC to T=225~K (D2) has increased, while the correlation length of the other two domains (D1 and D3) has decreased. This can be understood as a preferential population of one type of CO domain due to the electric field present at T$_{CO}$ leading to larger correlation length of these domains along the $c$-axis. 

Finally the sample was heated to T=225~K and cooled in an electric field (EFC) of 350~kV/m to T=180~K. The electric field was removed before the ND was recorded (Fig. \ref{fig_L}). Remarkably the intensity of all magnetic domains has increased significantly due to the electric field  present at the magnetic ordering temperature. At the same time the intensity of the diffuse rod, observed as the high background in Fig. \ref{fig_L}, has decreased. 

The correlation lengths after EFC to T=  180~K are largely the same as after EFC to T=225~K (Table \ref{table}). The additional magnetic intensity originates from magnetic domains of the same size, suggesting that the electric field assists in the nucleation of additional magnetic long range order.
Alternatively, the additional magnetic intensity can originate from an increase in the magnetic structure factor, as discussed next.

As now well understood, the CO in a single FeO$_5$ bipyramid layer consists of a 1:2 (or 2:1) ratio of Fe$^{2+}$ and  Fe$^{3+}$ ions. Such an arrangement is most stable for the Ising model on the triangular lattice. Because the magnetic moments exhibit a strong magnetic anisotropy along the $c$ axis as well as an AFM n.n. interaction \cite{wu_prl_2008}, a similar arrangement for the magnetic spins is expected, and indeed observed.\cite{ko_prl_2009} If the CO and MO are in-phase (in a single layer) all Fe$^{2+}$ moments are parallel and form one sublattice, and all Fe$^{3+}$ are antiparallel to the Fe$^{2+}$ moments, forming a second sublattice, see Fig. \ref{fig_layers}(a) bottom layer. However, the CO and MO can be out-of-phase which results in a more complex structure as indicated in the top layer of Fig. \ref{fig_layers}(a). Indeed it has been recently observed from magnetic x-ray dichroism studies \cite{ko_prl_2009, kuepper_prb_2009} that the majority Fe$^{2+}$ layer exhibits CO and MO in-phase while the minority Fe$^{2+}$ layer has CO and MO that are out-of-phase. This results in a bilayer where all the Fe$^{2+}$ moments are aligned parallel, while the Fe$^{3+}$ moment are either parallel or antiparallel (Fig \ref{fig_layers}). This magnetic arrangement in the bilayer has also been reported from ND studies by Ref. \cite{christianson_prl_2008}.

The spin only magnetic moments of the  Fe$^{2+}$ and  Fe$^{3+}$  ions are 4 $\mu_B$ and 5 $\mu_B$, respectively. Because these magnetic moments are distinct, the total moment of the minority and majority layer is different even if the magnetic sequence of up and down moments is the same. The total moment of a single layer is also different for MO that is in-phase or out-of-phase with the CO. Taking the spin only values and the magnetic sequence mentioned earlier, the above mentioned bilayer has $M^-$ = 3 $\mu_B$ per three iron ions in a majority Fe$^{2+}$ layer and $M^+$ = 4 $\mu_B$ per three iron ions in a minority Fe$^{2+}$ layer.

According to Hund's rules, Fe$^{3+}$ has five $3d$ electrons with spin up, and Fe$^{2+}$ has five $3d$ electrons with spin up and one with spin down. This latter $3d$ electron can jump from Fe$^{2+}$ to the Fe$^{3+}$ much easier if the spin down states of the latter are empty. This is the case for parallel alignment of the Fe$^{2+}$ and Fe$^{3+}$ moments. Thus electron hopping in the majority Fe$^{2+}$ layer is arrested because the Fe$^{2+}$  and Fe$^{3+}$ ions have opposite moments, while $3d$ electron hopping is promoted in the minority Fe$^{2+}$ layer because  some neighboring Fe$^{2+}$  and Fe$^{3+}$ ions have parallel moments. We also note that electron hopping between  Fe$^{2+}$  and Fe$^{3+}$ ions from one to the other layer is promoted, as their moments are parallel. This makes a convenient mechanism to switch between FE and AFE arrangements of the bilayers. As such the magnetism may assist the FE order to align with the electric field and enhance the electric polarization.

\begin{table}[t]
\vspace{-0.25cm}
\caption{The correlation lengths (in \AA) of the three magnetic domains along the $c$-axis recorded at T=180~K as function of electric field cooling procedure
(errors between brackets).
}
\center
\begin{tabular}{l ccc}
&{\bf D1} & {\bf D2} & {\bf D3}~~~~~\\
\hline\\
\vspace{-7mm}\\
\vspace{-5mm}\\
	ZEFC	         & 82(2) &75(3)	&84(1)\\
	EFC to T=225~K~~~~	&70(2)&92(1)&76(1)\\
	EFC to T=180~K	&85(1)&95(1)&75(1)\\
 \hline
\end{tabular}
\label{table}
\end{table}

We now turn to the interactions between the bilayers. Discussions are ongoing whether AFE or FE alignment of the bilayers is the ground state, suggesting these configurations are close in energy. AFM bilayer interactions have been reported by Ref. \cite{iida_jpsj_1993} while FM bilayer interaction was reported more recently.\cite{christianson_prl_2008} In the sample of interest in the present study the interaction is AFM, as witnessed by the half integer L superlattice. 

Fig \ref{fig_layers}(b,c) shows the AFM configuration for FE and AFE alignment of the bilayers. Because the total moment of the individual layers of the bilayer are $M^-$ = 3  $\mu_B$ and $M^+$ = 4 $\mu_B$ per three Fe ions respectively, the magnetic structures of Fig \ref{fig_layers}(b) and \ref{fig_layers}(c) are distinct. 
In the case that the MO is 3D while the CO is 2D, the order of $M^-$ and $M^+$ layers is random as drawn in Fig. \ref{fig_layers}(d). Such a magnetic arrangement gives rise to Bragg diffraction from layers with magnetic moments of 3 $\mu_B$  while 1 $\mu_B$ adds to the diffuse 2D rod. 
We estimate that a charge disordered state  (Fig. \ref{fig_layers}(d)) gives rise to $\sim$30\% less intensity at the L/2 reflections, compared to AFM order in a FE or AFE arrangement (Fig. \ref{fig_layers}(b) and (c)). 

Therefore, the observed increase in intensity at all L/2 reflections, combined with the decrease of intensity of the 2D rod, suggests that the electric field promotes the FE CO state over a 2D CO state. 
Furthermore, this FE state seems stabilized in the presence of AFM order as the measurements are performed after the electric field is removed.

This interpretation is corroborated  by the unchanged magnetic correlation length observed with or without EFC from 225~K to T=180~K. 
Once the CO bilayers are aligned by the electric field, the partial magnetic disorder is removed and the AFM order parameter increases while the domain size remains the same. We note the correlation length for D1 (Table \ref{table}) increased after EFC to T =180 K compared to EFC to T= 225 K. This possibly indicates that, in addition to the mechanism that we discuss, some regions may exhibit long range magnetic order after EFC only.

The intensity of the diffuse rod after EFC to T=180~K remains significant indicating that a large part of the sample shows 2D magnetic order. 
Given the diversity in magnetic responses reported for LuFe$_2$O$_4$ samples it is not unreasonable to suggest that FM and AFM regions can coexist
due to near degenerate magnetic ground states. \cite{degroot} We observe a strong reflection at ($\frac{1}{3} \frac{1}{3}$0) and a weaker one at ($\frac{1}{3} \frac{1}{3}$-3),  both present at T=180~K and T=225~K. In addition a broad peak appears at ($\frac{1}{3} \frac{1}{3}$-2) at T=225~K. These intensities are characteristic of FM coupling of the bilayers and they are independent of temperature and electric field cooling procedure.

It is tempting to attribute the increase in polarization observed at the magnetic ordering temperature due to the stabilization of long range CO by AFM regions. For a pyroelectric measurement the sample is cooled with electric field present at the magnetic ordering temperature T$_N$. This suggests that the degree of increase of electric polarization is related to the amount of 2D CO present at T$_N$, which is well below T$_{CO}$. 

The temperature dependence of the hopping frequency reported by Xu {\it et al} \cite{xu_prl_2008} shows charge fluctuations well below T$_{CO}$.
Nagano {\it et al.} \cite{nagano_prl_2007} calculate that spin order stabilizes the CO through frustration of the spin system which results in an entropy gain, and enhances the electric polarization below T$_N$. The present study suggests that the CO is stabilized with a well ordered spin system.

Wen {\it et al} \cite{wen_prb_2009} observe that the long range CO decreases when the sample is cooled in applied magnetic field of 7 T along [$\bar{1}$10], suggesting that there is correlation between the orientation of the magnetic moments and the formation of long range order, even in the paramagnetic state. This can be explained with the same phenomenon, parallel magnetic moments on neighboring Fe$^{2+}$ and Fe$^{3+}$ ions reduces the energy needed for the electron to hop from the Fe$^{2+}$ to the Fe$^{3+}$ ion.
We conclude that the alignment of the magnetic moments facilitates the hopping of electrons, promoting 3D FE order under applied electric field and promoting 2D CO without the presence of an electric field. 

We note that Wen {\it et al} \cite{wen_prb_2009} also reports that in the case of reduced long range CO, below T$_N$, the magnetic intensities at half-integer L are enhanced while those of integer L values have decreased. 
This is consistent with our observations and suggests the FM interactions between the bilayers occur in regions with 3D CO, while AFM interactions occur in regions with 2D CO.

In conclusion, we have demonstrated that the AFM order in LuFe$_2$O$_4$ can be changed with electric field cooling procedures. 
We propose that the increase of the electric polarization at the magnetic ordering temperature is due to regions that exhibit a transition from 2D CO with paramagnetic moments above T$_N$ to 3D CO with AFM moments below T$_N$.  While glassy orbitals \cite{mulders_prl_2009} and random magnetic moments give rise to a distribution of hopping integrals adding frustration to the charge order, the AFM order removes part of this frustration, causing a strong interaction between magnetic and charge degrees of freedom in this system.
These findings add to the understanding of the interaction between spin and charge, relevant for unconventional superconductivity, magneto-resistance, metal insulator transitions and other exotic electronic properties that arise in frustrated spin and charge systems.

The authors would like to thank Y. Wang  for pyroelectric current measurement, and acknowledge NCCR MaNEP Project and AINSE Ltd for providing financial assistance.

\end{document}